\documentclass[12pt]{article}
\usepackage{graphicx}
\title{Angular Spectra of Polarized Galactic Foregrounds}
\author{Jungyeon Cho, \& A. Lazarian}

\date{
Dept. of Astronomy, University of Wisconsin, 
    475 N. Charter St., Madison, WI53706; cho, lazarian@astro.wisc.edu
}

\begin{document}
\maketitle

\begin{abstract}

It is believed that magnetic field lines are twisted and bend by
turbulent motions in the Galaxy.
Therefore, both Galactic synchrotron emission and
thermal emission from dust reflects statistics of
Galactic turbulence.
Our simple model of Galactic turbulence, motivated by 
results of our simulations, predicts that
Galactic disk and halo exhibit different angular power spectra.
We show that observed angular spectra of synchrotron emission
are compatible with our model.
We also show that our model is compatible with
the angular spectra of star-light polarization for the Galactic disk.
Finally, we discuss how one can estimate polarized  
microwave emission from dust
in the Galactic halo using star-light polarimetry.

\end{abstract}


\section{Introduction}

Exciting possibilities of measuring spectrum of polarized 
cosmic microwave background (CMB) fluctuations
renewed the interest to the polarized fluctuations of
Galactic origin.
Synchrotron and dust are known to be the most important
sources of polarized foreground radiation.
Measurements of angular power spectra of such foregrounds are
of great interest.
So far, a large number of observations are available for 
angular power spectra of synchrotron emission 
and synchrotron polarization
for Galactic disk and halo (see papers in de Oliveira-Costa \& Tegmark 1999;
see also references listed in \S2 and papers in this volume).
Those measurements revealed a range of power-laws,
the origin of which has been addressed by Chepurnov (2002) and
Cho \& Lazarian (2002a; henceforth CL02).
The latter also addressed the origin of the observed
angular spectrum of starlight polarization 
(Fosalba et al. 2002; henceforth FLPT).

Interstellar medium is turbulent and Kolmogorov-type spectra
were reported on the scales from several AU to several kpc
(see Armstrong, Rickett, \& Spangler 1995; Lazarian \& Pogosyan 2000; 
Stanimirovic \& Lazarian 2001; Lazarian \& Esquivel 2003). 
Therefore it is natural to think of the 
turbulence as the origin of the fluctuations of the diffuse foreground
radiation. Interstellar medium is magnetized with magnetic
field making turbulence anisotropic. 
It may be argued that although the spectrum of
MHD turbulence exhibits scale-dependent anisotropy if
studied in the system of reference defined by the local
magnetic field\footnote{The anisotropy is present for Alfv\'{e}nic and slow
modes, while fast modes exhibit isotropy (see Cho \& Lazarian 2002b).} 
(Goldreich \& Sridhar 1995; Lithwick \&
Goldreich 2001; Cho \& Lazarian 2002b,2003ab),
in the observer's system of reference the spectrum will
show only moderate scale-independent anisotropy. Thus
from the observer's perspective Kolmogorov's description of
interstellar turbulence statistics is acceptable in spite of the fact that
magnetic field is dynamically important and even dominant (see discussion in 
Lazarian \& Pogosyan 2000; Cho, Lazarian, \& Vishniac 2003).  
  
In this paper, we claim
that MHD turbulence can explain observed
angular spectra of Galactic synchrotron emission  and
starlight polarization.
For this purpose, we use an analytical insight
obtained in Lazarian (1992, 1995ab) and numerical results
obtained in CL02..
We also discuss  how we can estimate polarized microwave dust emission
using star-light polarimetry.
This problem is of great importance in view of recent interest to
the foregrounds to the CMB polarization
(see also review by Lazarian in this volume).

\section{Galactic synchrotron emission}
\subsection{Summary of observations}
It is customary for
CMB studies to expand
the foreground intensity  over spherical harmonics $Y_{lm}$,
        $I(\theta, \phi)= \sum_{l,m}a_{lm}Y_{lm}(\theta,\phi)$, 
and write the spectrum in terms of  
        $C_l\equiv \sum_{m=-l}^{m=l} |a_{lm}|^2/(2l+1)$.
The measurements indicate that 
angular power spectrum ($C_l$) of the Galactic emission
follows power law ($C_l\propto l^{-\alpha}$)
(see FLPT and references in \S3).
The multipole moment $l$ is related to the angular scale 
$\theta$ on the sky as
$l\sim \pi/\theta^{radian}$
(or $l\sim 180^{\circ}/\theta^{\circ}$).
When the angular size of the observed sky patch
($\Delta \theta \times \Delta \theta$ in radian)
is small, 
$C_l$ is approximately the `energy' spectrum of fluctuations
(Bond \& Efstathiou 1987; Hobson \& Majueijo 1996; Seljak 1997).
It is this power spectrum expressed in terms of wavenumber
$k ~(\sim l\, \Delta \theta/\pi)$ 
that is usually dealt with in studies of astrophysical
turbulence (e.g. Stanimirovic \& Lazarian 2001). 

Recent statistical studies of {\it total} synchrotron intensity
include Haslam all-sky map at 408 MHz (Haslam et al.~1982) that
shows that the angular power index $\alpha$ ($C_l \propto l^{-\alpha}$) 
is in the 
range between
2.5 and 3 (Tegmark \& Efstathiou 1996; Bouchet, Gispert, \& Puget 1996).
Parkes southern Galactic plane survey (Duncan et al.~1997) 
at 2.4 GHz suggests shallower slope:
Giardino et al.~(2002) obtained $\alpha \sim 2.05$ after point source removal 
and Baccigalupi et al.~(2001) obtained $\alpha \sim 0.8$ to $2$.
On the other hand, Tucci et al.~(2000) obtained 
$\alpha \sim 1.4$ to $1.8$ and Bruscoli et al.~(2002) obtained 
$\alpha \sim 0.4$ to $2.2$ for the Galactic disk.
Using Rhodes/HartRAO data at 2326 MHz (Jonas, Baart, \& Nicolson 1998), 
Giardino et al.~(2001a) obtained 
$\alpha\sim 2.43$ for all-sky  data and $\alpha\sim 2.92$ for
high Galactic latitude regions with $|b|>20^{\circ}$. 
Giardino et al.~(2001b) obtained $\alpha\sim 3.15$ for 
high Galactic latitude regions with $|b|>20^{\circ}$ from
Reich \& Reich (1986) survey at 1420 MHz.
The rough tendency
that follow from these data is that $\alpha$ which is  close to $2$
for the Galactic plane gets steeper (to $\sim 3$) for higher latitudes 
(Fig.~\ref{fig_all}a).

Can we explain this tendency?
In the next section, we construct a simple model that can explain 
the rough tendency of observed angular spectra.

\subsection{Basic idea of our simple model}   \label{sect_idea}

\subsubsection{$C_l$ for small angle limit}
\begin{figure*}[!t]
\includegraphics[width=0.95\textwidth]{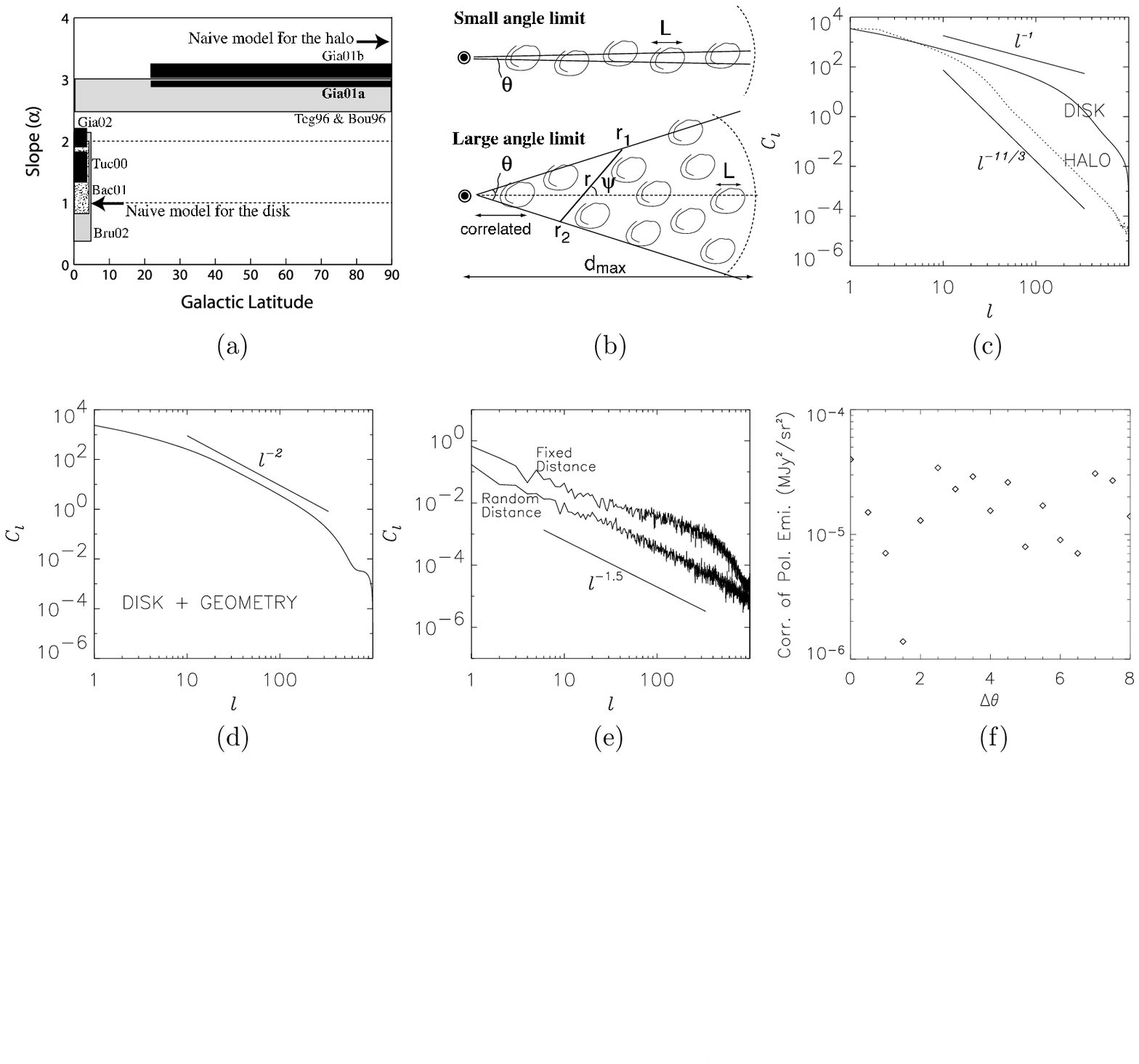}
\caption{
   (a) Comparison with observations.
              Widths of rectangles correspond to the Galactic latitude
              of the observed patches.
              Heights of rectangles denote either error bars or
              scatter of the slopes.
              ({\it Gia02}=Giardino et al.~2002; 
               {\it Bru02}=Bruscoli et al.~2002; etc.).
   (b)  {\it Upper plot:} 
              Small $\theta$ limit ($\theta <L/d_{max}$). The fluctuations
 along the entire length of the lines of sight are correlated.
              {\it Lower plot:}
              Large $\theta$ limit ($\theta >L/d_{max}$). 
              Only points close to the observer 
 are correlated.
   (c) Spectra for the disk 
             ($d_{max}$ = 10 kpc; turbulence size $L$ = 100 pc)
              and for the halo 
             ($d_{max}$ = 1 kpc; turbulence size $L$ = 100 pc).
   (d) Angular spectrum for the Galactic disk. 
              We considered the geometry of 
              the Galactic disk.
   (e) Angular spectra of degree of polarization.
      {\it Fixed Distance} means all stars are at the same distance of 10 kpc.
      {\it Random Distance} means random sampling of stars according 
      to equation
          (\ref{eq_13}).
          Zero points of $C(\theta)$ and $C_l$'s are arbitrary.
   (f) Angular correlation function for 
       polarized intensity at 1mm. From eq. (\ref{eq_Ipol_mm}).
       Figure (a)-(e) are from CL02.
         }
   \label{fig_all}
\end{figure*}

In this section we show that, when the angle between the lines of sight
is small (i.e.~$\theta < L/d_{max}$), the
angular spectrum $C_l$ has the same slope as the 3-dimensional
energy spectrum of turbulence.
Here $L$ is the typical size of the largest energy containing eddies,
which is sometime called as outer scale of turbulence or energy injection 
scale,
and
$d_{max}$ is the distance to the farthest eddies (see Fig.~\ref{fig_all}b).

To illustrate the problem consider the observations with lines of sight being
parallel. The observed intensity is the intensity summed
along the {\it line of sight}, $r_z$:
\begin{eqnarray}
 {I}_{2D}(r_x,r_y) & \equiv & \int d{r_z}~i_{3D}({\bf r})    \label{2d3d} \\
 & = & \int d{r_z}\ \int dk_x dk_y dk_z\ \hat{i}_{3D}({\bf k})\
  e^{i{\bf k}\cdot{\bf r}}.
\end{eqnarray}
Rearranging the order of summation and using 
$\int d{r_z}\ e^{ik_zr_z}=\delta(k_z)$, we get
\begin{equation}
  {I}_{2D}(r_x,r_y) = 
  \int d{k_x} dk_y ~\hat{i}_{3D}(k_x,k_y,0)   ~e^{ik_xr_x+ik_yr_y},
\end{equation}
which means Fourier transform of $ {I}_{2D}(r_x,r_y)$
is $\hat{i}_{3D}(k_x,k_y,0)$. As it was mentioned earlier 
for small patches of sky 
$C_l \sim |\hat{i}_{3D}(k_x,k_y,0)|^2$ with $l \sim k (\pi/\Delta \theta)$
and $k=(k_x^2+k_y^2)^{1/2}$.

The analysis of the geometry of crossing lines
of sight is more involved, but for power-law statistics it follows from
 Lazarian \& Shutenkov (1990) that if $|\hat{i}_{3D}|^2\propto k^{-m}$, 
then the `energy' spectrum
of  ${I}_{2D}(r_x,r_y)$ is also $k^{-m}$.  Therefore, we have 
\begin{equation}
   C_l \propto  |\hat{i}_{3D}(k_x,k_y,0)|^2 \propto l^{-m}.
\end{equation}
in the small $\theta$ limit.
For Kolmogorov turbulence ($|\hat{i}_{3D}|^2\propto k^{-11/3}$),
we expect
\begin{equation}
C_l \propto l^{-11/3}, \mbox{~~~if $\theta<L/d_{max}$.}
 \label{eq_5}
\end{equation}
Note that $l\sim \pi/\theta$.

\subsubsection{$C_l$ for large angle limit}

{}Following Lazarian \& Shutenkov (1990), we can show that
the angular correlation function for $\theta > L/d_{max}$ is given by
\begin{eqnarray}
    K(\theta) & = & \int \int dr_1 dr_2 ~{\cal K}( |{\bf r}_1-{\bf r}_2| ),
                          \nonumber
\\
              & = & \frac{1}{\sin{\theta}}
                    \int_0^{\infty}dr ~r {\cal K}(r) \int_0^{\pi/2}d\psi 
\sim \frac{const}{\theta},
\end{eqnarray}
where we change variables from $(r_1,r_2)$ to $(r,\psi)$, which is clear from
Fig.~1b. We accounted for the fact that
the Jacobian of the transformation is $r/\sin{\theta}$.
We can understand $1/\theta$ behavior qualitatively from Fig.~1a.
When the angle is large, 
the points along  of the lines of sight near the observer
are still correlated. These points extend from the observer
over the distance $\propto 1/\sin{(\theta/2)}$.

In the limit of $\theta\ll 1$ 
we get the angular power spectrum $C_l$ 
using Fourier transform:
\begin{eqnarray}
   C_l & \sim &  \int \int  K(\theta) 
                 e^{-i{\bf l}\cdot {\bf \theta}} d\theta_x d\theta_y  
\nonumber \\
       & \sim &  \int d\theta ~\theta J_0(l\theta) K(\theta) \propto   l^{-1},
   \label{eq_7}
\end{eqnarray}
where $\theta=(\theta_x^2+\theta_y^2)^{1/2}$, $J_0$ is the Bessel function, and
we use $K(\theta)\propto \theta^{-1}$.

In summary, for Kolmogorov turbulence, 
we expect from equations (\ref{eq_5}) and
(\ref{eq_7}) that
\begin{equation}
 C_l \propto \left\{ \begin{array}{ll} 
                         l^{-11/3}     & \mbox{if $l>l_{cr}$} \\
                         l^{-1}        & \mbox{if $l<l_{cr}$.}
                      \end{array}
              \right.
     \label{eq_1_11_3}
\end{equation}
which means that the power index $\alpha$ of $C_l$ is\footnote{ 
      Note that point sources would result in $\alpha \sim 0$.} 
$-1 \leq \alpha \leq -11/3$. 
The critical multipole moment $l_{cr} ~(\sim 3\,d_{max}/L)$ 
depends on the size of the large
turbulent eddies and on the direction of the observation. If in the
naive model we assume that turbulence is homogeneous along the lines
of sight and that $L\sim 100$ pc, corresponding to a typical size of the
supernova remnant, we get $l_{cr}\sim
300$ for the Galactic disk with $d_{max}\sim 10$~kpc.
For the synchrotron
halo, $d_{max}\sim 1$~kpc (see Smoot 1999) and we get $l_{cr}\sim
30$.     

\subsection{Numerical results}

Can this model really explain observed behavior of synchrotron 
angular spectrum?
To address the problem, we perform simple numerical calculations
for the Galactic disk and halo.
We obtain $C_l$ using the relation
\begin{eqnarray}
 & &K(\cos \theta) = \frac{1}{4\pi}\sum_{l} (2l+1)C_l P_l(\cos{\theta}), \\
 & &C_l = \frac{1}{2} \int P_l(\cos{\theta}) K(\cos \theta)\ d(\cos\theta).   
                                                             \label{c_l_leg}
\end{eqnarray}
We use Gauss-Legendre quadrature integration (see Szapudi et al.~2001 for its
application to CMB) to obtain $C_l$.
We numerically calculate the angular correlation function from
\begin{equation}
    K(\cos{\theta})=\int dr_1 \int dr_2 ~{\cal K}(|{\bf r}_1-{\bf r}_2|),
   \label{eq_ctheta}
\end{equation}
where $|{\bf r}_1-{\bf r}_2|=r_1^2+r_2^2-2r_1r_2 \cos{\theta}$ and
assume
that 
the spatial correlation function  ${\cal K}(r)$ follow Kolmogorov statistics:
\begin{equation}
 {\cal K}(r) \propto \left\{ \begin{array}{ll} 
                              {\cal K}_0-r^{2/3} & \mbox{if $r<L$} \\
                              0          & \mbox{if $r>L$,}
                      \end{array}
              \right. \label{c_r_3D}
\end{equation}
where ${\cal K}_0\sim L^{2/3}$ is a constant.
Fig.~\ref{fig_all}c 
illustrates the agreement of our calculations with the
theoretical expectations within the naive model of the disk and the
halo from the previous section.

\subsection{A more realistic model for Galactic disk}

The naive model discussed in the previous section
predicts $l^{-1}$ spectrum for the Galactic disk for $l < 300$
(see Fig.~\ref{fig_all}c and the discussion below eq.~(\ref{eq_1_11_3})).
However, as we discussed earlier,
observations show angular spectra close to $l^{-2}$.
In other words $2\leq \alpha \leq 3$ which differs from naive expectations
given by equation (\ref{eq_1_11_3}).


To make the spectrum closer to  observations we need
to consider more realistic models.
First, synchrotron emission is stronger in spiral arms, and therefore
we have more 
synchrotron emission coming from the nearby regions.
Second, if synchrotron disk component is sufficiently
thin,
then lines of sight are not equivalent and effectively nearby
disk component contributes more. Indeed,
when we observe regions with low Galactic latitude,
the effective line of sight varies with Galactic latitude.

Suppose we observe a region with $b=10^{\circ}$.
Then emission from $d=10$ kpc is substantially weaker than
that from $d=100$ pc, because the region with $d=10$ kpc is 
$d\sin{10^{\circ}}
\sim 1.7$ kpc above the Galactic plane and, therefore, has weak emission.
To incorporate the effect of finite thickness ($\sim100$ pc)
of the disk, we use the weighting function
$w(r)=100pc/\mbox{max}( 100pc, r \sin{10^{\circ}})$, which
gives more weight to closer distances.
The resulting angular power spectrum (Fig.~\ref{fig_all}d) 
shows a slope similar to -2.

For the halo, the simple model predicts that $C_l \propto l^{-11/3}$
for $l > 30$, but
observations provide somewhat less steep spectra.
Is this discrepancy very significant?
The spectrum of magnetic field is expected to be
shallower than $k^{-11/3}$ in the vicinity of the energy injection
scale and at the vicinity of the magnetic equipartition scale. 
The observed spectrum also gets shallower if $d_{max}$ gets larger.
For instance 
Beuermann, Kanbach, \& Berkhuijsen (1985) reported the existence of
thick radio halo that extends to more than $\sim 3~kpc$ near the Sun.
Finally, filamentary structures and point sources
can make the spectrum shallower as well. Further research should establish
the true reason for the discrepancy.

\section{MHD turbulence and Galactic Starlight Polarization}
Let us move to a different topic - starlight polarization.
In this section, we will illustrate that
MHD turbulence can explain the observed angular spectrum of 
starlight polarization.

Polarized radiation from dust is an important
component of Galactic foreground that strongly interferes with
intended CMB polarization measurements (see Lazarian \& Prunet 2001). 
FLPT attempted to predict the spectrum of the polarized
foreground from dust and 
obtained $C_l \sim l^{-1.5}$ for starlight polarization degree.
They used polarization data from $\sim$5500 stars. The sample
is strongly biased toward
nearby stars ($d<2$ kpc) within the Galactic plane.
This spectrum is different from those discussed in the previous sections.
To relate this spectrum to the underlying turbulence we should account
for the following facts: a) the observations are done for the disk component
of the Galaxy, b) the sampled stars are at different distances from
the observer with most of the stars close-by.

To deal with this problem we use numerical simulations again.
We first generate a three (i.e.~{\it x,y}, and {\it z}) 
components of magnetic field on a two-dimensional
plane ($4096 \times 4096$ grid points representing 
$10$ kpc $\times$ $10$ kpc), 
using the following Kolmogorov three-dimensional spectrum:
$E_{3D}(k)\propto k^{-11/3}$ if $k>k_0$,
where $k_0 \sim 1/100$pc. Our results show that the way 
how we continuously
extend the spectrum  for $k<k_0$ does not change our results.

To simulate the actual distribution of stars within the sample
used in FLPT, we scatter our emission sources
using the following
probability distribution function:
$  
   P(r) \propto e^{-r/1.5kpc}.    \label{eq_13}
$ 
The starlight polarization is due to the difference in absorption
cross section of non-spherical grains aligned with their longer axes
perpendicular to magnetic field (see a review by Lazarian 2003). In numerical
calculations we approximate the actual turbulent
magnetic field by a superposition
of the slabs with locally uniform magnetic field in each slap
and assume that the difference
in grain absorption parallel and perpendicular to magnetic field results
in the 10\% difference in the optical
depths $\tau_{\|}$ and $\tau_{\bot}$ for a slab. 
We calculate evolution of Stokes parameters of the starlight
within the slab and use the standard
transforms of Stokes parameters from one slab to another 
(Dolginov, Gnedin, \& Silantev 1996; see similar 
expressions in Martin 1972).

We show the result in Fig.~\ref{fig_all}e.
For comparison, we also calculate the degree of polarization assuming
all stars are at the same distance of $\sim10$ kpc.
The result shows that, for a mixture of nearby and faraway stars,
the slope steepens and gets very close to the observed one, i.e.
$-1.5$.

\section{Estimation of Polarized Emission from Dust}
One of the possible ways to estimate the polarized radiation 
{}from dust at the microwave
range is to measure star-light polarization and use
the standard formulae relating polarization at different wavelength.
The technique can be traced back to FLPT.
In this section, we estimate polarized diffuse emission by
dust in the high Galactic latitude halo.
For our practical data handling, see Hildebrand et al. (1999).
We use the data of optical depth (in fact, $E(B-V)$) and 
the degree of polarization by absorption $P_{abs,optical}$ 
for stars in the halo (provided by Terry Jones).

When the optical depth is small, we have the following relation (see, for
example, Hildebrand et al. 2000):
\begin{equation}
  P_{em,optical} \approx -P_{abs,optical}/\tau,
\end{equation}
where $P_{em,optical}$ is the degree of polarization by emission and
$\tau$ is the optical depth (at optical wavelengths).
We obtain polarization by emission at 1mm ($P_{em,mm}$) using the relation
\begin{equation}
  P_{em,mm}= P_{em,optical}
         \left[ \frac{ C_{max}-C_{min} }{ C_{max}-C_{min} } \right]_{mm}
      /  \left[ \frac{ C_{max}-C_{min} }{ C_{max}-C_{min} } \right]_{optical},
  \label{eq_pem_mm}
\end{equation}
where $C$'s are cross sections
that depend on the geometrical shape (see, for example, the discussion in 
Hildebrand et al. 1999; see also Draine \& Lee 1984)
and dielectric function $\epsilon$ 
(see Draine 1985) of 
grains.
We assume grains are 2:1 oblate spheroids.

We can obtain the emission at 1mm ($I_{mm}$) 
{}from that at 100 $\mu m$ ($I_{100\mu m}$):
\begin{equation}
     I_{mm} = I_{100\mu m}  ( 1mm/100\mu m )^{-\beta},
\end{equation}
where we use $\beta\sim 1.7$.
For now, we assume that  $I_{100\mu m}\approx 0.85$ MJy/sr 
(Boulanger \& Perault 1988; see also Draine \& Lazarian 1998).
{}From this, we obtain $I_{mm} \approx 0.017$ MJy/sr.
In the future, we can use an actual $100\mu m$ map to get
$I_{mm}$.

{}From $P_{em,mm}$ and $I_{mm}$, we can easily obtain
the polarized intensity at 1mm ($I_{pol,mm}$):
\begin{equation}
  I_{pol,mm}= P_{em,mm} I_{mm}.
\label{eq_Ipol_mm}
\end{equation}
Fig. \ref{fig_all}f, obtained this way, shows $I_{pol,mm}$.

Last step will be the estimation of the angular spectrum $C_l$.
{}From CL02 (see also \S\ref{sect_idea}), we obtain
\begin{equation}
   C_l \approx \left\{ \begin{array}{ll}
                                      Al^{-1} & \mbox{ if $l<30$ } \\
                           30^{8/3}Al^{-11/3} & \mbox{ if $l>30$, }
                       \end{array}
                       \right.
\end{equation}
where A is a constant that is determined by
$
    \sum l(l+1)C_l \approx 10^{-5},
$
where $10^{-5}$ is the rms fluctuation (see y-axis of Fig. \ref{fig_all}f).
Calculation yields $A=10^{-5}/30^2$. Therefore,
\begin{equation}
      C_l \approx \left\{ \begin{array}{ll}
                         10^{-7}l^{-1} ~~(MJy^2/sr^2) & \mbox{ if $l<30$ } \\
                      10^{-4}l^{-11/3} ~~(MJy^2/sr^2) & \mbox{ if $l>30$. }
                       \end{array}
                       \right.
\end{equation}

When we prefer $(\mu K)$ to $(MJy/sr)$, we can use the conversion factor
 $c_* \approx (1/x^2) (10^4 \mu K/(MJy/sr))$,
where $x\approx \nu/56.8$ GHz (Tegmark et al.~2000).
For 1mm, $x\approx 5$ and $c_* \approx 400$.
$C_l c_*^2$ gives
\begin{equation}
      C_l \approx \left\{ \begin{array}{ll}
                         10^{-2}l^{-1} ~~(\mu K^2) & \mbox{ if $l<30$ } \\
                          10 l^{-11/3} ~~(\mu K^2) & \mbox{ if $l>30$. }
                       \end{array}
                       \right.
\end{equation}
The particular normalization coefficients and the spectral index may vary
{}from one region to another (see FLPT; CL02).

We can summarize the procedure as follows:
\begin{equation}
     \left.  \begin{array}{crl}
    P_{abs,optical}  & \rightarrow P_{em,optical} & \rightarrow P_{em,mm}  \\
      & I_{100\mu m}&\rightarrow I_{mm}  
                       \end{array}
                       \right\}\rightarrow C_l \mbox{~(at 1 mm)}
\end{equation}

\section{Summary}
In this paper we have addressed the origin of spatial
fluctuations
of Galactic diffuse emission.
We have shown that MHD turbulence with Kolmogorov spectrum
can qualitatively explain the observed
properties of synchrotron emission and starlight polarization.
The variety of measured spatial spectra of synchrotron emission
 can be accounted for by the
inhomogeneous distribution of emissivity along the line of sight
arising from the structure of the Galactic disk and halo.
Similarly, MHD turbulence plus inhomogeneous distribution of stars 
can explain the reported scaling of starlight polarization
statistics. Although these complications do not allow to predict
the exact slope of the measured fluctuations, our interpretation
allows a valuable qualitative insight on what sort of change
is to be expected.
 
We have shown
that one can estimate polarized diffuse micro-wave emission by dust
using star-light.
Our preliminary result shows that $\delta T \sim [l(l+1)C_l]^{1/2}
\sim O(1) \mu K$ for high Galactic latitude regions.

Evidently more systematic studies are required.
Those studies will not only give insight into how to
separate CMB from foregrounds, but also
would provide valuable information on
the structure of interstellar medium and the sources/energy
injection scales of interstellar turbulence.
(See also review by Lazarian, this volume.)

\vspace{0.7cm}
\noindent
{\bf Acknowledgments}
We acknowledge the support of NSF Grant AST-0125544.


\begin{thebibliography}{9}
\bibitem{} Armstrong, J., Rickett, B., \& Spangler, S. 1995, ApJ, 443, 209

\bibitem{} Baccigalupi, C., Burigana, C., Perrotta, F., De Zotti, G.,
             La Porta, L., Maino, D., Maris, M., \& Paladini, R. 2001
             A\&A, 372, 8

\bibitem{} Beuermann, K., Kanbach, G., \& Berkhuijsen, E. 1985,
             A\&A, 153, 17
\bibitem{} Bond, J.R. \& Efstathiou, G. 1987, MNRAS, 226, 655

\bibitem{} Bouchet, F.R., Gispert, R., \& Puget, J.L. 1996,
             in {\it Unveiling the Cosmic Infrared Background},
             AIP Conf.~Proc.~348, ed.~E.~Dwek (Baltimore: AIP), 225

\bibitem{}Boulanger, F. \& Perault, M. 1988, ApJ, 330, 964
\bibitem{} Bruscoli, M., Tucci, M., Natale, V.,
             Carretti, E., Fabbri, R., Sbarra, C., \& Cortiglioni, S. 
             2002, NewA, 7, 171


\bibitem{} Chepurnov, A. V. 2002, astro-ph/0206407 
           (Astron.Astrophys.Trans. 17 (1999) 281-300)

\bibitem{} Cho,J. \& Lazarian, A. 2002a, ApJL, 575, L63 (CL02)
\bibitem{} Cho, J. \& Lazarian, A. 2002b, Phy.~Rev.~Lett., 88, 245001

\bibitem{} Cho, J., Lazarian, A., \& Vishniac, E.~T. 2003, 
             in
      {\it Turbulence and Magnetic Fields in Astrophysics},
      eds. E. Falgarone \& T. Passot (Springer LNP), p56
      (astro-ph/0205286) 


\bibitem{} de Oliveira-Costa, A. \& Tegmark, M. 1999, 
             {\it Microwave Foregrounds}, ASP Conf.~Ser.~181, 
             (San Francisco: ASP)


\bibitem{} Dolginov, A.~Z., Gnedin, Iu.~N., \& Silantev, N.~A. 1996,
             {\it Propagation and Polarization of Radiation in Cosmic Media},
             (Gordon \& Breech)

\bibitem{} Draine, B. 1985, ApJS, 57, 587
\bibitem{} Draine, B. \& Lazarian, A. 1998, ApJL, 494, L19
\bibitem{} Draine, B. \& Lee, H. 1984, ApJ, 285, 89


\bibitem{} Duncan, A. R., Haynes, R. F., Jones, K. L., \& Stewart, R. T.
             1997, MNRAS, 291, 279
\bibitem{} Fosalba, P., Lazarian, A., Prunet, S., \& Tauber, J.A. 2002
             ApJ, 564, 762 (FLPT)

\bibitem{} Giardino, G., Banday, A.J., Fosalba, P., G\'{o}rski, K.M.,
             Jonas, J.L., O'Mullane, W., \& Tauber, J. 2001a, A\&A, 371, 708

\bibitem{} Giardino, G., Banday, A. J., Bennett, K., Fosalba, P.,
             G\'{o}rski, K. M., O'Mullane, W., Tauber, J., \& Vuerli, C.
             2001b, in {\it Mining the Sky},
             ed.~ A.~J. Banday et al. (Springer-Verlag), 458
             (astro-ph/0011084)
             

\bibitem{} Giardino, G., Banday, A.J., G\'{o}rski, K.M., Bennett, K.,
             Jonas, J.L., \& Tauber, J. 2002, A\&A, 387, 82

\bibitem{} Goldreich, P. \& Sridhar, H. 1995, ApJ, 438, 763 

\bibitem{} Haslam, C.G.T., Stoffel, H., Salter, C.J., \& Wilson, W.E.
             1982, A\&AS, 47, 1

\bibitem{} Hildebrand, R. et al. 2000, PASP, 112, 1215
\bibitem{} Hildebrand, R. et al. 1999, ApJ, 516, 834

\bibitem{} Hobson, M.P. \& Magueijo, J. 1996, MNRAS, 283, 1133

\bibitem{} Jonas, J. L., Baart, E. E., \& Nicolson, G. D.
             1998, MNRAS, 297, 977

\bibitem{} Lazarian, A. 1992, Astron.~\& Astrophys.~Trans., 3, 33
\bibitem{} Lazarian, A. 1995a, Ph.~D.~Thesis (Univ. of Cambridge, UK)
\bibitem{} Lazarian, A. 1995b, A\&A, 293, 507
\bibitem{} Lazarian, A. 2003, 
                J. of Quantitative Spectroscopy \& Rad. Transf., 79-80, 881

\bibitem{} Lazarian, A. \& Esquivel, A. 2003, ApJL, in press (astro-ph/0304007)
\bibitem{} Lazarian, A. \& Pogosyan, D. 2000, ApJ, 537, 720
\bibitem{} Lazarian, A. \& Prunet, S. 2001, in
             {\it Astrophysical Polarized Backgrounds}, AIP Conf.~Proc.~609, 
             ed.~S.~Cecchini et al.
             (Melville: AIP), 32
\bibitem{} Lazarian, A. \& Shutenkov, V.~P. 1990, PAZh, 16,
690 (translated Sov.~Astron.~Lett., 16, 297)



\bibitem{} Lithwick, Y. \& Goldreich, P. 2001, ApJ, 562, 279

\bibitem{} Martin, P.~G. 1972, MNRAS, 159, 179

\bibitem{} Reich, P. \& Reich, W. 1986, A\&AS, 63, 205

\bibitem{} Seljak, U. 1997, ApJ, 482, 6
\bibitem{} Smoot, G.~F. 1999, in {\it Microwave Foregrounds}, 
             ASP Conf.~Ser.~181, 
             ed. A.~de Oliveira-Costa \& M. Tegmark (San Francisco: ASP), 61


\bibitem{} Stanimirovic, S., \& Lazarian, A. 2001, ApJ, 551, L53
\bibitem{} Szapudi, I., Prunet, S., Pogosyan, D., Szalay, A. S., \&
             Bond, J. R. 2001, ApJ, 548, L115


\bibitem{} Tegmark, M. \& Efstathiou, G. 1996, MNRAS, 281, 1297
\bibitem{} Tegmark, M., Eisenstein, D.~J., Hu, W., de Oliveira-Costa, A.
             2000, ApJ, 530, 133

\bibitem{} Tucci, M., Carretti, E., Cecchini, S., Fabbri, R.,
             Orsini, M., \& Pierpaoli, E. 2000, NewA, 5, 181

\end{thebibliography}
\end{document}